\documentclass[twocolumn,preprintnumbers,amsmath,amssymb]{revtex4}

\usepackage{graphicx}
\usepackage{dcolumn}
\usepackage{bm}

\def\smh{\hspace*{0.3cm}}

\begin{document}

\title{Exactly Soluble Models for Surface Partition}

\author{K. A. Bugaev$^{1,2}$ and J. B. Elliott$^{2}$}
\affiliation{$^1$Bogolyubov Institute for Theoretical Physics,
Kiev, Ukraine\\
%
%
$^2$Lawrence Berkeley National Laboratory, Berkeley, CA 94720, USA
}

\date{\today}
\begin{abstract}
The surface partition of large clusters is studied analytically 
within a frame-work of the ``Hills and Dales Model''.
Three 
formulations
are solved exactly by using the Laplace-Fourier 
transformation method. 
In the limit of  small amplitude  deformations,
the ``Hills and Dales Model''  gives upper and lower
bounds for the surface entropy coefficient of large clusters.  
A comparison with the 2- and 3-dimensional Ising model surface entropy coefficients is made.
\end{abstract}

\maketitle


\section{Introduction}

{
The surface entropy of large clusters was introduced by Fisher in 
his droplet model (FDM) \cite{Fisher:67}. 
From the study of the combinatorics of  clusters,
Fisher postulated that the leading contribution to the  surface entropy is  
proportional to the surface $S$, i.e. ${\omega S}$ (in dimensionless units). 
The coefficient  
$\omega$ is a ratio  of the surface energy  coefficient  
$\sigma_{\rm o} (T_c)$ per one constituent  taken at  
the critical temperature $T_c$.
The  surface entropy  was studied  recently
in our paper \cite{Bugaev:04b}.  
There we developed the ``Hills and Dales Model'' (HDM) which is a statistical model of 
surface deformations that  obeys the volume conservation of considered clusters. 
Using the novel mathematical method, the Laplace-Fourier transform \cite{Bugaev:04},
we were able to find  the grand canonical surface partition (GCSP) of the HDM analytically. 
For
vanishing deformations we 
obtained the upper limit for the surface entropy 
coefficient $\omega$ of large  clusters  to be $\omega \approx 1.06009$ 
(in dimensionless units),
i.e.  
about 6 \% larger than  Fisher's postulate. 
 
In the grand canonical formulation the cluster volume is conserved in average, but 
to apply the HDM  to  small and finite clusters it is necessary to consider an exact
volume conservation using the canonical formulation.  
Therefore, in present paper we  consider the canonical formalism for the HDM
and obtain the infimum for the surface entropy of finite and large clusters.   
For the limit of  vanishing deformations  we also  introduce the 
{\it semi-grand canonical ensemble} which occupies an intermediate 
place between the grand canonical and  canonical surface ensembles. 
With the help of the Laplace-Fourier transform technique \cite{Bugaev:04}
the canonical surface partition (CSP)  and the  semi-canonical surface partition (SGCSP)  
are  evaluated exactly  for any volume of cluster. 
For large clusters
the leading contribution  and its corrections   are found analytically
for the CSP and SGCSP.  
The obtained values for the $\omega$-coefficient  are compared with the corresponding 
values for  2- and 3-dimensional Ising models for different lattice geometries.   
It is shown that $\omega$ values of all 2- and 3-dimensional Ising models 
lie between the supremum and infimum of the HDM. 

The paper is organized as follows. In  Sect. II we formulate three ensembles for surface
deformations within the HDM frame work and solve them analytically 
by the Laplace-Fourier transform technique.
Sect. III is devoted to the analysis of the isochoric ensemble singularities and to the
derivation of the upper estimates of the surface entropy coefficient.  The lower estimates for
the surface entropy coefficients  are found and compared to the corresponding 2- and 3-dimensional
Ising   lattice values in   Sect. IV. The conclusions are formulated in Sect. V.  
}

\vspace*{-0.25cm}

\section{Hills and Dales Model}

\vspace*{-0.25cm}

The HDM is  a  statistical model of surface deformations. 
We impose a necessary constraint that the deformations should conserve the total volume of 
the cluster of $A$-constituents. 
As in our  previous paper \cite{Bugaev:04b},
the  main interest is focused on 
the deformations of vanishing amplitudes. 
This is sufficient to find both an absolute supremum and absolute infimum for the
$\omega$-coefficient of the HDM.
In this case   the shape  of the deformation
cannot be important to our result, so  we can
choose the regular one.
For this reason  we shall consider  cylindrical deformations of positive height $h_k>0$ (hills)
and negative height $-h_k$ (dales), with  $k$-constituents at the base.
For simplicity it is assumed that the top (bottom) of the hill (dale) has the same shape as
the surface of the original  cluster of $A$-constituents.
We also  assume that:
(i) the statistical weight of deformations $\exp\left( - \sigma_{\rm o} |\Delta S_k|/s_1 /T  \right) $
is given  by the Boltzmann factor due to the  change of the surface $|\Delta S_k|$ in units of
the surface per  constituent $s_1$;
(ii) all hills of heights $h_k \le H_k$ ($H_k$ is the maximal height of
the hill with $k$-constituents at the base)
have the same probability $d h_k/ H_k$ besides the statistical one;
(iii) assumptions (i) and (ii) are valid for the dales.

These assumptions are not too restrictive and allow us to simplify the analysis and
find the one-particle statistical partition of
the deformation of the $k$-constituent base as a convolution of two probabilities
{ discussed above:}
\begin{equation}\label{one}
z_k^{\pm} \equiv \hspace*{-0.15cm} \int\limits_0^{\pm H_k} \hspace*{-0.15cm} \frac{ d h_k}{ \pm H_k}\,
{\textstyle e^{ - \frac{\sigma_{\rm o} P_k |h_k| }{T s_1} } }
= T s_1 
\frac{\left[1 - {\textstyle e^{ - \frac{\sigma_{\rm o} P_k H_k}{ T s_1} } } \right] }{\sigma_{\rm o} P_k H_k }\,,
\end{equation}
where upper (lower) sign corresponds to hills (dales). Here $P_k$ is the perimeter of the cylinder base. 
Our next step is 
to find a geometrical partition (degeneracy factor) or the number of ways to place
the center of a  given  deformation on the surface of the $A$-constituent cluster 
which  is occupied
by the set of $\{n_l^\pm  = 0, 1, 2,...\}$  deformations of the $l$-constituents base. 

 For the grand canonical surface partition (GCSP)
the desired geometrical partition
can be given  in the  excluded volume  approximation \cite{Bugaev:04b}
as
\begin{equation}\label{two}
{\cal G}_{gc} = 
{\textstyle \left[ S_A - \sum\limits_{k = 1}^{K_{max} } k\, (n_k^+ ~ + ~ n_k^-) \, s_1 \right] s_1^{-1} } \,, 
\end{equation}
where $s_1 k$ is the area  occupied by the deformation of $k$-constituent base ($k = 1, 2,...$), 
$ S_A$
is the  full surface of the cluster,
and $K_{max} (S_A) $ is the $A$-dependent size of the maximal allowed base on the cluster.   
It is clearly seen now that 
the first multiplier in the right hand side (r.h.s.) of (\ref{two}) corresponds to the available surface
to place the center of each of $\{n_k^\pm \}$ deformations that exist on the cluster surface.
It is necessary to impose the condition ${\cal G}_{gc} \ge 0$ which ensures that the deformations
do not overlap.
Equation (\ref{two}) is the van der Waals excluded volume  
approximation usually used in statistical mechanics at low particle densities 
\cite{Bondorf:95,Goren:81, Bugaev:00, Bugaev:01, Reuter:01} 
and can  be derived for  objects of different sizes in 
the spirit of Ref.  \cite{Zeeb:02}.

According to Eq. (\ref{one}) the statistical partition for the hill with a $k$-constituent base matches
 that of the dale, i.e.   $z^+_k = z^-_k$, and, therefore,  
the GCSP
\begin{eqnarray}
\hspace*{-0.50cm}& &Z_{gc}(S_A)= 
\nonumber \\ 
\label{three}
\hspace*{-0.50cm}& &\sum\limits_{\{n_k^\pm = 0 \}}^\infty \hspace*{-0.10cm} \left[ \prod_{k=1}^{K_{max} }
\frac{ \left[ z_k^+ {\cal G}_{gc} \right]}{n^+_k!}^{n^+_k} \frac{ \left[ z_k^- {\cal G}_{gc} \right]}{n^-_k!}^{n^-_k}\right]
\Theta(s_1 {\cal G}_{gc} )\, 
\end{eqnarray}
corresponds to the conserved (on average) volume of the cluster because the probabilities 
of hill and dale of the same base are identical. 
The $\Theta(s_1 {\cal G}_{gc})$-function in (\ref{three}) ensures that only configurations
with positive value of the free surface of cluster are taken into account,  
but
this makes the calculation of the GCSP very difficult. 

For small and finite clusters we have to impose a more strict constraint of
the exact volume conservation of cluster.
This can be done within the canonical ensemble assuming that the number of the hills $n_k^+$ of 
the $k$-constituent base is always identical to the number of corresponding dales, i.e.
$n_k^- \equiv n_k^+ \equiv n_k$. Then the canonical geometrical partition can be cast as follows 
\begin{equation}\label{nfour}
{\cal G}_c =
{\textstyle \left[ S_A - 2 \sum\limits_{k = 1}^{K_{max} } k\, n_k \, s_1 \right] (2 s_1)^{-1} } \,,
\end{equation}
where the factor two in the denominator of the right hand side (r.h.s.) of (\ref{nfour})
accounts for the fact that it is necessary to place simultaneously the centers of 
two $k$-constituent base deformations (hill and dale) out of $2 n_k$ on the surface of cluster. 
Using the geometrical partition (\ref{nfour}), one can write  the partition function of 
canonical ensemble, the CSP,  
as follows
\begin{equation} \label{nfive}
Z_c(S_A)= \hspace*{-0.10cm} \sum\limits_{\{n_k = 0 \}}^\infty \hspace*{-0.10cm} \left[ \prod_{k=1}^{ K_{max} }
\frac{ \left[ z_k^+ z_k^-  {\cal G}_c \right]}{n_k!}^{n_k} \right]
\Theta(2 s_1 {\cal G}_c)\,.
\end{equation}
As in the case of the GCSP,
the $\Theta(2 s_1 {\cal G}_c)$-function in the CSP ensures that only configurations
with positive value of the free surface of cluster are accounted for,
but this constraint  makes the calculation of the partition (\ref{nfive}) very difficult.

An additional problem in evaluating 
the partitions (\ref{three}) and (\ref{nfive}) appears due to 
the explicit $S_A$ dependence
of the maximal base of  deformations via $K_{max} (S_A)$
because in this case 
the standard method to deal with
the excluded volume partitions, the usual Laplace transform  
\cite{Goren:81, Bugaev:00, Bugaev:01, Reuter:01} in $S_A$,
cannot be applied.  
However, as shown in \cite{Bugaev:04b} the GCSP (\ref{three}) can be solved analytically 
with the help of the 
Laplace-Fourier technique \cite{Bugaev:04}. The latter employs the identity 
\begin{equation}\label{nsix}
G (S_A) =
%
\int\limits_{-\infty}^{+\infty} d \xi~ \int\limits_{-\infty}^{+\infty}
  \frac{d \eta}{\sqrt{2 \pi}} ~
{\textstyle e^{ i \eta (S_A - \xi) } } ~ G(\xi)\,,
\end{equation}
which is based on the Fourier representation of the Dirac $\delta$-function.
Similar to the GCSP,  
the representation (\ref{nsix}) allows us to decouple the additional  
$S_A$-dependence in $K_{max} (S_A)$  of the CSP and reduce it to the exponential one,
which can  be integrated by the Laplace transform \cite{Bugaev:04, Bugaev:04b}
\begin{eqnarray} \label{nseven}
&&\hspace*{-0.5cm}{\cal Z}_c(\lambda)~\equiv ~\int_0^{\infty}dS_A~{\textstyle e^{-\lambda S_A}}
~{ Z}_c(S_A) = \nonumber\\
&&\hspace*{-0.5cm}\int_0^{\infty}\hspace*{-0.2cm}dS^{\prime}
\int\limits_{-\infty}^{+\infty} d \xi~ \int\limits_{-\infty}^{+\infty}
\frac{d \eta}{\sqrt{2 \pi}} ~ { \textstyle e^{ i \eta (S^\prime - \xi) - \lambda S^{\prime} } } 
\times
\nonumber \\
&&\hspace*{-0.5cm}
\sum\limits_{\{n_k = 0 \}}^\infty
\hspace*{-0.1cm} \left[\prod_{k=1}^{K_{max}(\xi)} \right.
\left.
\frac{ \left[ z_k^+ z_k^- S^{\prime} {\textstyle e^{ 2 k\,s_1(i \eta -\lambda ) } } \right]}
{n_k!~(2~ s_1)^{n_k} }^{n_k} 
\right]
\Theta(S^{\prime}) =
\nonumber \\
&&\hspace*{-0.5cm}\int_0^{\infty}\hspace*{-0.2cm}dS^{\prime}
\int\limits_{-\infty}^{+\infty} d \xi~ \int\limits_{-\infty}^{+\infty}
 \frac{d \eta}{\sqrt{2 \pi}} ~ { \textstyle e^{ i \eta (S^\prime - \xi) - \lambda S^{\prime}
+ S^\prime {\cal F}_c(\xi, \lambda - i \eta) } }\,.
\end{eqnarray}
After changing the integration variable 
$S_A \rightarrow S^{\prime} = S_A - 2 \sum\limits_{k = 1}^{K_{max}(\xi) } k\, n_k \, s_1 $,
the constraint of $\Theta$-function has disappeared.
Next all $n_k$ were summed independently leading to the exponential function.
Now the integration over $S^{\prime}$ in (\ref{nseven})
can be  done giving 
 the canonical isochoric partition
\begin{equation}\label{neight}
\hspace*{-0.4cm}{\cal Z}_c(\lambda) = \int\limits_{-\infty}^{+\infty} \hspace*{-0.1cm} d \xi
\int\limits_{-\infty}^{+\infty} \hspace*{-0.1cm}
\frac{d \eta}{\sqrt{2 \pi}} ~
\frac{ \textstyle e^{ - i \eta \xi } }{{\textstyle \lambda - i\eta ~-~{\cal F}_c(\xi,\lambda - i\eta)}}~,
\end{equation}

\vspace*{-0.3cm}

\noindent
where the function ${\cal F}_c(\xi,\tilde\lambda)$ is defined as follows
\begin{equation}\label{nnine}
{\cal F}_c(\xi,\tilde\lambda) = \sum\limits_{k=1}^{ K_{max}(\xi) } 
 \frac{ z_k^+\, z_k^-}{2\,s_1} 
~e^{ - 2\, k\,s_1\tilde\lambda  }\,.
\end{equation}
The representation (\ref{neight}) is generic - it is also valid for the GCSP, if 
the canonical function (\ref{nnine}) is replaced  by  the grand canonical one 
\begin{equation}\label{nten}
{\cal F}_{gc}(\xi,\tilde\lambda) = \sum\limits_{k=1}^{ K_{max}(\xi) }
\left[  \frac{ z_k^+}{s_1} + \frac{ z_k^-}{s_1} \right]
~e^{ - k\,s_1\tilde\lambda  }\,.
\end{equation}

Before 
making  the inverse Laplace transform and studying the structure of singularities 
of the functions (\ref{nnine}) and (\ref{nten}),
it is necessary to discuss one more ensemble for the surface deformations 
which hereafter will be called as
the {\it semi-grand canonical surface partition}. 

This ensemble occupies an intermediate position between the canonical and grand canonical 
formulations. It corresponds to the case, when the hills and dales 
of the same base are considered to be indistinguishable. For that  we would like to explore the
fact that according to (\ref{one})  the statistical probabilities of hills and dales
of the same base are equal. Then for the infinitesimally small amplitudes of deformations  
the volume conservation constraint is fulfilled trivially.  
For finite amplitudes of deformations this ensemble corresponds to the case of
clusters of  intermediate size,
when  its surface is sufficiently filled with deformations that it is
impossible to add  one more  deformation of large base in order  
to conserve the cluster volume, but it is  possible  to place many deformations of
a much smaller (or even smallest) base. 

For this case the geometrical factor reads as  
\begin{equation}\label{neleven}
{\cal G}_{sg} =
{\textstyle \left[ S_A -  \sum\limits_{k = 1}^{K_{max} } k\, n_k \, s_1 \right]  s_1^{-1} } \,,
\end{equation}
and the SGCSP has the following form  
\begin{equation} \label{ntwelve}
Z_{sg}(S_A)= \hspace*{-0.10cm} \sum\limits_{\{n_k = 0 \}}^\infty 
\hspace*{-0.10cm} \left[ \prod_{k=1}^{ K_{max} }
\frac{ \left[ z_k^+   {\cal G}_{sg} \right]}{n_k!}^{n_k} \right]
\Theta( s_1 {\cal G}_{sg})\,.
\end{equation}
It is easy to show that using the Laplace-Fourier transform technique \cite{Bugaev:04} 
the SGCSP (\ref{ntwelve}) can be transformed into the generic representation (\ref{neight})  
for the  function 
\begin{equation}\label{nthirteen}
{\cal F}_{sg}(\xi,\tilde\lambda) = \sum\limits_{k=1}^{ K_{max}(\xi) }
\frac{ z_k^+}{s_1} ~e^{ - k\,s_1\tilde\lambda  }\,.
\end{equation}
By construction 
the equations (\ref{ntwelve}) and (\ref{nthirteen}) are less fundamental than
the corresponding  grand canonical and canonical functions.

\vspace{-0.25cm}

\section{Analysis of  Singularities}

\vspace*{-0.2cm}

To study the structure of  singularities of the isochoric partition (\ref{neight}),
it is necessary to make the inverse Laplace transform ($\alpha \in \{gc, sg, c\}$):
%
\begin{eqnarray*}
&&\hspace*{-1.8cm} Z_\alpha (S_A)~ =
\int\limits_{\chi - i\infty}^{\chi + i\infty}
\frac{ d \lambda}{2 \pi i} ~  {\cal Z}_\alpha (\lambda)~ e^{\textstyle   \lambda \, S_A } =
\end{eqnarray*}
%
%
\begin{eqnarray}\label{nfourteen}
&&\hspace*{-0.8cm}\int\limits_{-\infty}^{+\infty} \hspace*{-0.1cm} d \xi
\int\limits_{-\infty}^{+\infty} \hspace*{-0.3cm}\frac{d \eta}{\sqrt{2 \pi}}
\hspace*{-0.1cm} \int\limits_{\chi - i\infty}^{\chi + i\infty}
\hspace*{-0.1cm} \frac{ d \lambda}{2 \pi i}~
\frac{\textstyle e^{ \lambda \, S_A - i \eta \xi } }{i
{\textstyle \lambda - i\eta ~-~{\cal F}_\alpha (\xi,\lambda - i\eta)}}~=
\nonumber \\
&&\hspace*{-0.8cm}\int\limits_{-\infty}^{+\infty} \hspace*{-0.1cm} d \xi
\int\limits_{-\infty}^{+\infty} \hspace*{-0.3cm}\frac{d \eta}{\sqrt{2 \pi}}
\,{\textstyle e^{  i \eta (S_A - \xi)  } } \hspace*{-0.1cm} \sum_{\{\tilde\lambda_n\}}
%
%
{\textstyle 
e^{\textstyle  \tilde\lambda_n\, S_A }\hspace*{-0.1cm}
\left[1 - \frac{\partial {\cal F}_\alpha
(\xi,\tilde\lambda_n)}{\partial \tilde\lambda_n} \right]^{-1} \hspace*{-0.2cm},
}
\end{eqnarray}
where the contour  integral in $\lambda$ 
is reduced to the sum over the residues of all singular points
$ \lambda = \tilde\lambda_n + i \eta$ with $n = 0, 1, 2,..$, since this  contour 
in the complex $\lambda$-plane  obeys the
inequality $\chi > \max(Re \{  \tilde\lambda_n \})$.
Now all integrations in (\ref{nfourteen}) can be done, and all three surface partitions 
($\alpha \in \{gc, sg, c\}$)
can be written as 
\begin{equation}\label{nfifteen}
 Z_\alpha (S_A) ~ = \sum_{\{\tilde\lambda_n\}}
e^{\textstyle  \tilde\lambda_n\, S_A }
{\textstyle
\left[1 - \frac{\partial {\cal F}_\alpha
(S_A,\tilde\lambda_n)}{\partial \tilde\lambda_n} \right]^{-1} } \,,
\end{equation}
i.e. the double integral in (\ref{nfourteen}) simply  
reduces to the substitution   $\xi \rightarrow S_A$ in
the sum over singularities.
This remarkable answer for all three surface partitions   
is a partial example  of the general theorem on the Laplace-Fourier transformation
properties proved in \cite{Bugaev:04}.

The simple poles in (\ref{nfourteen}) are defined by the condition 
$\tilde\lambda_n~ = ~{\cal F}_\alpha (S_A,\tilde\lambda_n)$ and 
 for each ensemble
the latter can be cast
as a system of two coupled transcendental equations
\begin{eqnarray}\label{nsixteen}
&&\hspace*{-0.2cm} R_n^\alpha = ~  \sum\limits_{k=1}^{K_{max}(S_A) } \phi^\alpha_k  
~{\textstyle e^{- k\,R_n^\alpha} } \cos(I_n^\alpha  k)\,,
\\
\label{nseventeen}
&&\hspace*{-0.2cm} I_n^\alpha = -  \sum\limits_{k=1}^{K_{max}(S_A) } \phi^\alpha_k 
~{\textstyle e^{-k\,R_n^\alpha} } \sin(I_n^\alpha  k)\,, 
\end{eqnarray}
for dimensionless variables defined as $R_n^\alpha = s_1 Re(\tilde\lambda_n)$ and 
$I_n^\alpha = s_1 Im(\tilde\lambda_n)$ for the GCSP and SGCSP, 
and as $ R_n^c = 2 s_1 Re(\tilde\lambda_n)$ and
$I_n^c = 2 s_1 Im(\tilde\lambda_n)$ for the CSP.
Here the function $\phi^\alpha_k$ is  given by the expression   
\begin{eqnarray}\label{neighteen}
\phi^\alpha_k = 
\left\{
\begin{tabular}{ll}
\vspace{0.1cm} $   z_k^+ + z_k^-  $\,, &  for $\alpha = gc$\,, \\
$  z_k^+ z_k^-  $\,, & for $\alpha = c$\,, \\
$ z_k^+  $\,, & for $\alpha = sg$\,.
\end{tabular}
\right.
\end{eqnarray}

To this point  Eqs. (\ref{nsixteen}), (\ref{nseventeen}) and (\ref{neighteen})
are  general and can be used for
particular models which specify the height of hills and depth of dales. 
But 
it is possible to give both the upper and lower estimates 
for all three  partition functions of  large clusters,
and even to estimate corrections for finite and small clusters. 
For the upper estimate 
let us consider 
the real root $(R_0^\alpha ; I_0^\alpha = 0)$ of these equations.  
It is sufficient to  consider the  limit $K_{max}(S_A) \rightarrow \infty$,
because for $I_n^\alpha = I_0^\alpha = 0$ 
the r.h.s.  of (\ref{nsixteen}) is a monotonously increasing function of $K_{max}(S_A)$.
Since $z_k^+ = z_k^- $ are the monotonously decreasing functions of $H_k$, the maximal value of 
the r.h.s. of (\ref{nsixteen})  corresponds to the limit of infinitesimally small amplitudes 
of deformations, $H_k \rightarrow 0 \Rightarrow z_k^+ = z_k^- = 1 $ .
Under these conditions Eq. (\ref{nseventeen}) for 
$I_n^\alpha = I_0^\alpha = 0$ becomes an identity and  
Eq. (\ref{nsixteen}) acquires the form 
\begin{equation}\label{nnineteen}
R_0^\alpha =  B^\alpha \sum\limits_{k=1}^{\infty } 
~{\textstyle e^{ - k\,R_0^\alpha} }  = B^\alpha \left[ e^{ R_0^\alpha } - 1 \right]^{-1} \,,
\end{equation}
where $B^{gc} = 2$ and $ B^{c} = B^{sg} = 1$. 
Therefore, the real roots of (\ref{nsixteen}) and 
the corresponding  surface entropy coefficients $\omega^{\alpha}_U$
are as follows 
\begin{eqnarray}\label{ntwenty}
R_0^\alpha =
\left\{
\begin{tabular}{rl}
\vspace{0.1cm} $ \omega^{gc}_U = \max\{\omega^{gc}\} \approx  1.060090 $\,, &\,  for $\alpha = gc$\,, \\
$2 \omega^{c}_U = 2 \max\{\omega^{c}\} \approx   0.806466 $\,, &\, for $\alpha = c$\,, \\
$\omega^{sg}_U = \max\{\omega^{sg}\} \approx  0.806466 $\,, &\, for $\alpha = sg$\,.
\end{tabular}
\right.
\end{eqnarray}
The results  (\ref{ntwenty}) correspond to the upper estimate for the
surface partitions because
for $I_n^\alpha \neq 0$ defined by (\ref{nseventeen}) the inequality $\cos(I_n^\alpha k) \le 1$
cannot become the equality for all values of $k$ simultaneously. Then 
it follows that the real root of (\ref{nsixteen}) obeys the inequality $R_0^\alpha > R_{n > 0}^\alpha$.  
The last result means that in the limit of infinite cluster, $S_A \rightarrow \infty$, 
all surface partitions (\ref{nfifteen}) are
represented by the farthest right singularity among all simple poles $\{\tilde\lambda_n\}$ 
\begin{equation}\label{ntwone}
\max\{ Z_\alpha (S_A) \} ~ \rightarrow
\frac{e^{\textstyle  \omega^\alpha_U \frac{S_A}{s_1} } }{1 + 
\frac{R_0^\alpha ( R_0^\alpha + B^\alpha) }{B^\alpha} } 
=  g_\alpha\, e^{\textstyle  \omega^\alpha_U \frac{S_A}{s_1} }\,,
\end{equation}
where the geometrical degeneracy prefactor $g_\alpha$ is defined as follows:
$g_{gc} \approx 0.38139$ and $g_{c} = g_{sg} \approx 0.407025$.
Thus, the geometrical factor of the leading term for all three models is
practically the same. 

Remarkably    
the result (\ref{ntwone}) is  model independent.
This is a consequence of 
the  vanishing deformations limit
in which all model specific parameters vanish.
The second remarkable
fact is that Eq. (\ref{ntwone})
is valid for any self-non-intersecting surfaces of cluster.
This is so because both  {\it the shape and dimensionality  of the cluster under consideration 
do not enter in our equations explicitly.}  For our analysis of the HDM  
surface partitions 
it  was sufficient to require  that the cluster surface 
together with deformations is a regular surface
without  self-intersections.
Therefore, for vanishing deformations the latter means that Eq. (\ref{ntwone})
should be   valid for any self-non-intersecting surfaces.  



%
%
\begin{figure}[ht]
\includegraphics[width=8.6cm,height=8.6cm]{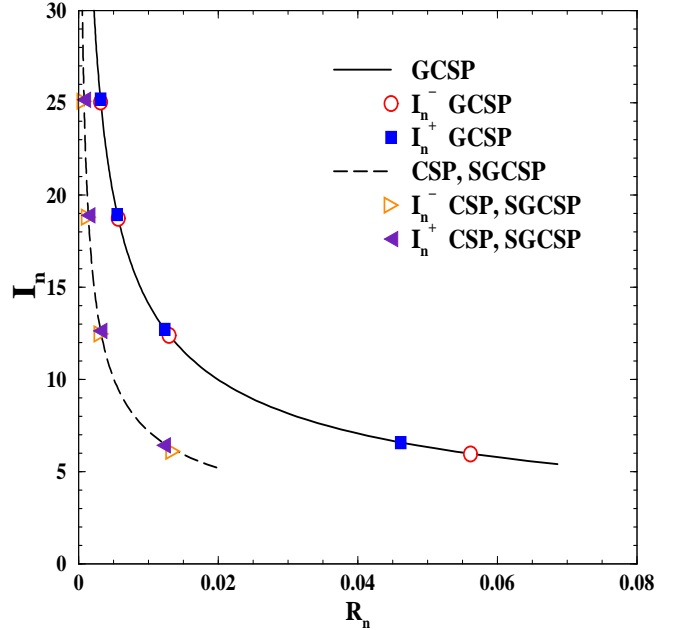}
\caption{
The first quadrant of the complex plane $s_1\tilde\lambda_n \equiv R_n + i I_n$ shows the roots
of the system of Eqs. (\ref{nsixteen}) and (\ref{nseventeen}).
The symbols   represent the two branches $I_n^-$ and $I_n^+$ of the roots
for the upper estimate of three surface partitions.
The curve is defined by the approximation given by
(\ref{ntwtwo}) and  (\ref{ntwthree}) (see text for more details).
}
  \label{fig1}
\end{figure}


For  large, but finite clusters it is necessary to take into account not only the farthest
right singularity $\tilde\lambda_0 $ in (\ref{nfifteen}),  but all other roots with 
positive real part  
$R_{n>0}^\alpha > 0$. 
The analysis presented in Appendix A shows that besides the 
opposite signs there are two branches of solutions, $I^{\alpha \,+}_{n}$ and $I^{\alpha \,-}_{n}$, 
for the same $n  \ge 1$ value: 
\begin{equation}
\label{ntwtwo}
|I^{\alpha \, \pm}_{n}|  \approx  2 \pi n  \pm \frac{B^\alpha}{2 \pi n}\,, 
%
\end{equation}
\vspace*{-0.3cm}
\begin{equation}
\label{ntwthree}
R_{n}^\alpha  \approx  \frac{(B^\alpha)^2}{8 \pi^2 n^2}\,.
\end{equation}
The exact solutions $(R_n^\alpha ; I^{\alpha\, \pm}_n)$ for $ n \ge 1$ 
which have the largest real part  are shown in Fig. 1
together with the curve  parametrized by functions $I_x^{\alpha\,+}$ and $R_x^\alpha$
taken  from Eqs. (\ref{ntwtwo}) and  (\ref{ntwthree}),
respectively.
From Eq. (\ref{ntwthree}) and Fig. 1 it is clear that for the GCSP 
the  largest real part $R_1^{gc} \approx 0.0582$ is 
about 18 times smaller than $R_0^{gc}$, 
whereas for the CSP and SGCSP the  real part $R_1^{c} = R_1^{sg}$
of the first most right complex root of Eqs. (\ref{nsixteen}) and (\ref{nseventeen})
is about 63.6 times smaller than $R_0^{c} = R_0^{sg}$.  
Therefore,  for a cluster of a few constituents   the correction 
to the leading term (\ref{ntwone}) is exponentially small for all considered partitions.
Using the approximations (\ref{ntwtwo}) and  (\ref{ntwthree}),
for $n > 2$ one can  estimate  
the upper limit of
the $(R_n^\alpha ; I^{\alpha\, \pm}_n)$ root contribution into Eq. (\ref{nfifteen})
\begin{equation}\label{ntwfour}
\hspace*{-0.2cm}\biggl| {\textstyle  e^{\tilde\lambda_n\, S_A} 
\left[1 - \frac{\partial {\cal F}_\alpha (S_A,\tilde\lambda_n)}{\partial \tilde\lambda_n} 
\right]^{-1} }\biggr|  \le
e^{\textstyle  \frac{(B^\alpha)^2\,S_A}{ 8 \pi^2 n^2 s_1} }\, /\, (2 \pi^2 n^2) \,. 
\end{equation}
This result shows that 
for all three considered partition
the total contribution of all complex poles in (\ref{nfifteen})
is negligibly small compared to the leading term (\ref{ntwone}) 
for a cluster of a few constituents or more.


\section{The Surface Entropy Coefficients}

\vspace{-0.25cm}

To  complete our analysis of the limit of vanishing  deformations 
we would like to find the lower estimate for the GCSP, CSP and SGCSP  for large clusters.  
This estimate corresponds to the absence of all other deformations except for  
those of smallest base. In other words, one has to substitute $K_{max} (S_A) = 1$  
in all corresponding expressions.
Then equations (\ref{nsixteen}) and (\ref{nseventeen}), respectively
become
\begin{eqnarray}\label{ntwfive}
&&\hspace*{-0.2cm} R_n^\alpha = ~   \phi^\alpha_1
~{\textstyle e^{- R_n^\alpha} } \cos(I_n^\alpha )\,,
\\
\label{ntwsix}
&&\hspace*{-0.2cm} I_n^\alpha = -  \phi^\alpha_1
~{\textstyle e^{- R_n^\alpha} } \sin(I_n^\alpha )\,.
\end{eqnarray}
Similar to the previous consideration, 
the leading term of the lower estimate for the surface partitions (\ref{nfifteen}) is given by the 
real root $(R_0^\alpha ; I^{\alpha}_0 = 0)$ of the system  (\ref{ntwfive}) and (\ref{ntwsix})
\begin{eqnarray}\label{ntwseven}
R_0^\alpha =
\left\{
\begin{tabular}{rl}
\vspace{0.1cm} $ \omega^{gc}_L = \min\{\omega^{gc}\} \approx  0.852606 $\,, &\,  for $\alpha = gc$\,, \\
$2 \omega^{c}_L = 2 \min\{\omega^{c}\} \approx   0.567143 $\,, &\, for $\alpha = c$\,, \\
$\omega^{sg}_L = \min\{\omega^{sg}\} \approx  0.567143 $\,, &\, for $\alpha = sg$\,.
\end{tabular}
\right.
\end{eqnarray}
%

%
%
\begin{figure}[ht]
\includegraphics[width=8.6cm,height=8.6cm]{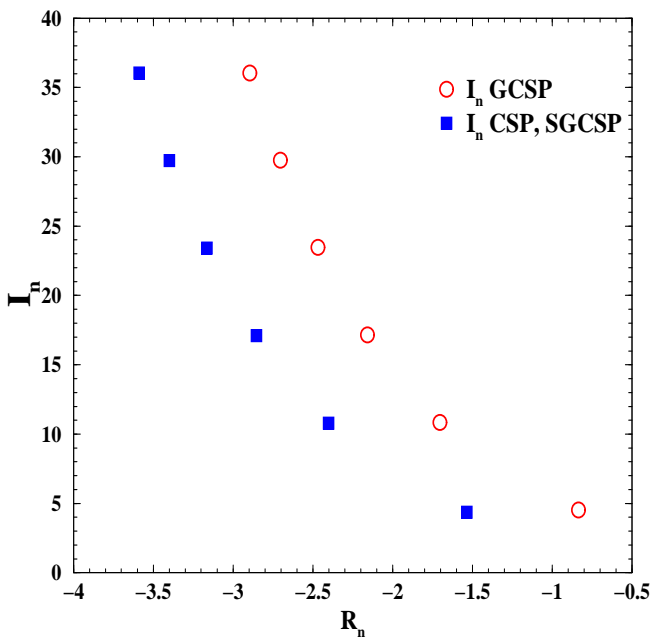}
\caption{
The  second quadrant of the complex plane $s_1\tilde\lambda_n \equiv R_n + i I_n$ shows the complex roots
of the system of  Eqs. (\ref{ntwfive}) and (\ref{ntwsix}) with the largest real parts.
The circles and squares  represent  the roots
for the lower estimate of the GCSP and CSP(SGCSP), respectively.
}
  \label{fig2}
\end{figure}


Again, as  in case of upper estimates one can show that the real root 
$(R_0^\alpha ; I^{\alpha}_0 = 0)$  approximates well the lower estimate for the  partition function
for a system of a few constituents.  
In fact, each of three surface partitions  has only a single root with positive real part which coincides with $(R_0^\alpha ; I^{\alpha}_0 = 0)$.  In Fig. 2  a few  complex roots  of Eqs.  (\ref{ntwfive}) and (\ref{ntwsix}) with the largest real parts are 
shown. Since all these roots have negative real part, they generate an exponentially small contribution
to the  lower estimate of surface partition for a system of a few constituents.


\begin{center}
\begin{tabular}{|c|c|c|}\hline
 & & \\
\smh Partition \smh &  \smh $\max\{\omega^\alpha\}$ \smh  & \smh  $\min\{\omega^\alpha\}$ \smh  \\
 & & \\
\hline
\hline
GCSP   & 1.060090   & 0.852606 \\
\hline
SGCSP  & 0.806466 & 0.567143 \\
\hline
CSP  & 0.403233  & 0.283572  \\
\hline
\end{tabular}
\\
%

\end{center}

\noindent
 Table I.
{\small
The  maximal and minimal values of the $\omega$-coefficient for
three statistical partitions of the HDM.  
}

\vspace*{0.5cm}

The $\omega$-coefficients  for  upper and lower estimates of  all three surface  partitions are summarized  in the Table I.
A comparison with the corresponding coefficient for liquids should be made with care because
various contributions to the surface tension, i.e., eigen surface tension
of the liquid  drop, the  geometrical degeneracy factor (surface partition), and the part  
induced by interaction between clusters, are not exactly known.  Therefore, even
the  linear  temperature dependence of the surface tension $\sigma(T) = \sigma_{\rm o} (T_c -  T)/T_c$ due to Fisher \cite{Fisher:67}
applied to a nuclear  liquid ( $ \sigma_{\rm o} \approx 18$ MeV;  $T_c \approx  18$ MeV \cite{Bondorf:95}) may be used to estimate the $\omega$-coefficient, if  both
the eigen surface tension and the interaction induced one are non-increasing functions of 
temperature. Under these assumptions one can get  the following inequality for 
nuclear liquid
\begin{equation}\label{ntweight}
 \omega_{nucl}~ \le ~ 1  ~ <~  \omega^{gc}_U~ = ~1.060090\,, 
\end{equation}
i.e. the upper estimate for the GCSP, indeed, provides  the upper limit for surface partition
of nuclear matter.

A similar analysis for real liquids is  difficult  because of  complicated 
temperature dependence of the surface tension.
Therefore, we would like to compare the $\omega$-coefficients from Table I with 
the $\omega$-coefficients for  the large spin clusters of various 2- and 3-dimensional
Ising models, which are listed in the Tables II and III, respectively \cite{Fisher:69}. 
Such a comparison can be made because  the surface entropy  of large spin clusters 
on the Ising lattices are similar to the considered surface  partitions (\ref{nfifteen}) \cite{Fisher:69}.

The $\omega$-coefficient for the d-dimensional  Ising model is defined as 
the energy   $ 2 J$ required to flip a given spin interacting with its  $q$-neighbors  to opposite direction 
per  $(d-1)$-dimensional surface  divided by the value of critical temperature
\begin{equation}\label{ntwnine}
\omega_{Lat}  = \frac{q \,\,J}{T_c \,d}\,.
\end{equation}
Here $q$ is the coordination number for the lattice, and $J$  denotes the coupling constant of the model. A comparison of the Tables I - III shows that  all lattice $\omega_{Lat}$-coefficients, indeed,
lie between the  upper estimates for canonical and grand canonical surface partitions
\begin{equation}\label{nthirty}
0.403233 ~= ~\omega^{c}_U  ~< ~ \omega_{Lat} ~ <~  \omega^{gc}_U~ = ~1.060090\,, 
\end{equation}
i.e. $\omega^{c}_U$ and $\omega^{gc}_U$ are the infimum and supremum for 2- and 3-dimensional
Ising models, respectively.


\begin{center}
\begin{tabular}{|c|c|}\hline
 &  \\
\smh Lattice type \smh &  \smh $ \omega_{Lat} = \frac{\sigma}{T_c} $ \smh   \\
 &  \\
\hline
\hline
\smh Honeycomb  \smh &\smh  0.987718\smh \\
\hline
Kagome  & 0.933132  \\
\hline
Square  & 0.881374   \\
\hline
Triangular  & 0.823960  \\
\hline
Diamond  & 0.739640   \\
\hline
\end{tabular}
\\
%

\end{center}

\noindent
 Table II.
{\small
The values of the $\omega_{Lat}$-coefficient for various
2-dimensional Ising models.
For more details see the text.
}

\vspace*{0.3cm}


\begin{center}
\begin{tabular}{|c|c|}\hline
 &  \\
\smh Lattice type \smh &  \smh $ \omega_{Lat} = \frac{\sigma}{T_c} $ \smh   \\
 &  \\
\hline
\hline
\smh Simple cubic  \smh &\smh  0.44342\smh \\
\hline
\smh Body-centered cubic \smh & 0.41989  \\
\hline
Face-centered cubic  & 0.40840   \\
\hline
\end{tabular}
\\
%

\end{center}

\noindent
 Table III.
{\small
The values of the $\omega_{Lat}$-coefficient for various
3-dimensional Ising models.
}

\vspace*{0.4cm}

The HDM partitions do not have an explicit dependence on the dimension of the surface,
but a comparison of
the HDM and Ising model $\omega$-coefficients shows that the HDM ensembles 
seem to  posses   some sort of  internal 
dimension: the GCSP is close to honeycomb, kagome or square lattices, whereas the SGCSP
is similar to triangular and diamond lattices, and  the $\max\{\omega\}$ of the  CSP is closer to the 3-dimensional
Ising models.  In some cases the agreement with the lattice data  is remarkable - $\omega^{gc}_L$
coincides with the arithmetical  average of 
the $  \omega$-coefficients for  square and triangular lattices up
to a fifth digit, but in most cases the values agree within a few per cent. 
The latter is not surprising because the HDM estimates  the surface entropy of 
a single cluster, whereas on the lattice the spin  clusters  do interact with each other and this, of course,  changes the surface tension and, consequently, affects the value of  critical temperature.
It is remarkable that so oversimplified estimates of the surface partitions for
a single large cluster 
reasonably  approximate the  $  \omega$-coefficients for 2- and 3-dimensional Ising models.
 
It would be interesting to check whether the lower estimate of the CSP $\omega_L^c \approx 0.283572$
is an infimum for the Ising lattices of  higher dimensions $d > 3$. 
If this is the case, then we can give an upper limit for the critical temperature of those lattices
using Eq. (\ref{ntwnine})
\begin{equation}\label{nthone}
\frac{T_c}{J}~ \le~ \frac{q}{\omega_L^c\, d}~ \approx~  3.5264~ \frac{q}{d} \,. 
\end{equation}
On the other hand,
the lower estimate
for the critical temperature of  Ising lattices, $ \frac{T_c}{J}~ \ge~ \frac{q}{\omega_U^{gc}\, d}$,
is provided by 
the supremum of the $\omega$-coefficients of surface partitions.




\section{Conclusions}

\vspace{-0.25cm}

We have formulated the grand canonical and canonical  partitions 
of surface deformations in the frame work of the HDM.  Both partitions conserve  the volume 
of the deformed cluster
and take into account all  surface deformations  with non-negative value of the free surface
of this cluster. The grand canonical surface partition conserves the cluster volume  in average,
whereas in canonical formulation it is conserved exactly.
These partitions  are solved exactly for  an  arbitrary (finite or infinite) 
size of largest deformation by the Laplace-Fourier transformation  
technique and the the general analytical  expression (\ref{nfifteen})
for these partitions 
in terms of the set  of isochoric ensemble  singularities 
are derived.

Similarly  we introduced and solved a special ensemble, a semi-grand canonical partition, which  obeys all
constraints discussed above in the limit of  vanishing deformations and occupies an intermediate
place between the grand canonical and canonical ensembles. 

Then we considered the limit of vanishing deformations for all three surface partitions, and obtained
the upper and lower estimates for the surface entropy for each of these partitions. 
A comparison of obtained $\omega$-coefficients for surface partitions with the corresponding
coefficients for the large spin clusters of  2- and 3-dimensional Ising models shows 
that the upper estimate of the GCSP is a supremum, whereas the upper estimate of the CSP
is an infimum for considered lattices. The question of the Ising  lattice  $\omega$-coefficients for
higher dimensions is discussed.  

The developed  formalism is rather general and, therefore, may be
applied to the  surface deformations of any kind of clusters,
if the underlying mechanism of the surface deformations is given.

{\bf  Acknowledgments.}
This work was supported by the US Department of Energy.

\vspace*{0.75cm}

\section{Appendix A}

For  large, but finite clusters it is necessary to take into account not only the farthest
right singularity $\tilde\lambda_0 $ in (\ref{nfifteen}),  
but all other roots 
of Eqs. (\ref{nsixteen}) and (\ref{nseventeen})
which have 
positive real part
$R_{n>0}^\alpha > 0$. In this case for each $R_{n>0}^\alpha$ there are two roots $\pm I_n^\alpha$
of (\ref{nseventeen}) because
the partition function (\ref{nfifteen})
 is real by definition.
The roots of Eqs. (\ref{nsixteen}) and (\ref{nseventeen}) with largest real part are insensitive
to the large values of $K_{max} (S_A)$, therefore, it is sufficient to  keep
$K_{max} (S_A) \rightarrow \infty$.
Then for limit of vanishing amplitude of deformations Eqs. (\ref{nsixteen}) and
(\ref{nseventeen}) can be, respectively, rewritten as
\begin{eqnarray}\label{Aone}
&&\hspace*{-0.2cm}\frac{B^\alpha R_n^\alpha}{(R_n^\alpha)^2 + (I_n^\alpha)^2} =
~~{\textstyle e^{R_n^\alpha} } \cos(I_n^\alpha ) - 1\,,
\\
\label{Atwo}
&&\hspace*{-0.2cm}\frac{B^\alpha I_n^\alpha}{(R_n^\alpha)^2 + (I_n^\alpha)^2} = -
~{\textstyle e^{R_n^\alpha} } \sin(I_n^\alpha )\,.
\end{eqnarray}
After some algebra the system of (\ref{Aone}) and (\ref{Atwo})
can be  reduced to a single equation for $R_n^\alpha$
\begin{eqnarray}\label{Athree}
&&\hspace*{-0.5cm}
\cos\left( {\textstyle \left[ 
\frac{ B^\alpha (B^\alpha + 2 R_n^\alpha)}{e^{2 R_n^\alpha} - 1} - (R_n^\alpha)^2 \right]^{1/2} } \right) = 
\nonumber \\
&&\hspace*{2.0cm}\cosh R_n^\alpha - \frac{B^\alpha}{B^\alpha + 2~ R_n^\alpha} \sinh R_n^\alpha\,,
\end{eqnarray}
and  the  quadrature
$I_n^\alpha = \sqrt{\frac{B^\alpha (B^\alpha + 2 R_n^\alpha)}{e^{2 R_n^\alpha}- 1} - (R_n^\alpha)^2 }$.
The analysis shows that besides the
opposite signs there are two branches of solutions, $I^{\alpha\,+}_{n}$ and $I^{\alpha\,-}_{n}$,
for the same $n  \ge 1$ value.
Expanding both sides of (\ref{Athree}) for $R_n^\alpha \ll 1$ and keeping the leading terms, 
one obtains (\ref{ntwtwo}) and  (\ref{ntwthree}).
In Fig. 1 this approximation is compared  with a few  exact solutions 
$(R_n ; I^\pm_n)$ for $ n \ge 1$ which have the largest real part.
%
%
%
%
%
%
%
%
%


\end{document}